\documentclass[a4paper, 11pt]{article}
\date{March  2009}
\usepackage{epsfig}

\newcommand{\be}{\begin{equation}}
\newcommand{\ee}{\end{equation}}
\newcommand{\ba}{\begin{eqnarray}}
\newcommand{\ea}{\end{eqnarray}}
\newcommand{\bi}{\begin{itemize}}
\newcommand{\ei}{\end{itemize}}

\renewcommand{\>}{\rangle}

\newcommand{\la}{\label}

\def\a{\alpha}      \def\b{\beta}   \def\g{\gamma}      \def\G{\Gamma}
\def\d{\delta}      \def\D{\Delta}  \def\e{\varepsilon}
          \def\l{\lambda}     
\def\m{\mu}                     
             
\def\r{\varrho}       
\def\t{\tau}          
\def\x{\xi}              
\def\w{\omega}      \def\W{\Omega}  
  \def\OO{{\cal O}}
\def\pa{\partial} 

\newcommand\Tr{\mbox{Tr}}
\def\sumint{\hbox{$\sum$}\!\!\!\!\!\!\int}

\begin{document}
\begin{titlepage}
\begin{flushright}
CPT-P036-2009\\
ITFA-12-2009\\

\end{flushright}
\begin{centering}
\vfill

 \vspace*{2.0cm}
{\bf \Large Spatial 't Hooft loop % to cubic order
 in  hot SUSY theories at weak coupling}

\vspace{2.0cm}
\centerline{{\bf Chris~P.~Korthals Altes}}
\centerline{Centre Physique Th\'eorique au CNRS}
\centerline{Case 907, Campus de Luminy}
\centerline{F-13288 Marseille, France, and}

\vspace{0.5cm}

%\centerline{\bf Harvey~B.~Meyer}
\centerline{Institute Theoretical Physics}
\centerline{Valkeniersstraat 65}
\centerline{ 1018 XE, Amsterdam, the Netherlands}
%\vspace{0.1cm}
\centerline{altes@cpt.univ-mrs.fr}

\vspace*{2.0cm}

\end{centering}
\vfill
\centerline{\bf Abstract } 
\vspace{0.1cm}
\noindent
The spatial 't Hooft  loop  measures the colour electric flux in SU(N)/Z(N) gauge theory. It is a closed loop of Dirac Z(N) flux and has strength k=1, 2,.., N-1.  It is analyzed for generic k and small gauge coupling in the high temperature phase of {\cal  N} =1, 2 and 4 SUSY theory  up and including cubic order.  In one loop order no qualitative difference with  gluodynamics shows. However the two loop order shows a   logarithmic 
divergence appear in the centre of the wall. This is because the gluinos 
become bosonic in the centre and acquire long wavelength excitations. 
 We discuss the cure for this divergence, due to a  gluino zero mode. The physics of Casimir scaling
 is explained. The cubic order can be obtained from gluodynamics by simple rescalings of the Debye masses and the   Casimir parameter $k/N_c$.

\vfill
 
 \end{titlepage}

\setcounter{footnote}{0}

%%%%%%%%%%%%%%%
\section{Introduction\la{sec:intro}}

The physics at RHIC is that of a strongly interacting
plasma.  In this regime QCD can only be studied on the lattice like  the smallness of the shear viscosity over entropy ratio (for a review see ref.~\cite{pis2008}).   Many questions are still beyond the reach of simulations, like the fast rate in going to equilibrium.
This is one of the reasons to contemplate the AdS/CFT connection and its non-equilibrium ramifications~\cite{jdeboer}. The CFT 
is four dimensional high temperature ${\cal N}=4$ SUSY, and can be studied in strong coupling by 
the calculating in the five dimensional bulk theory in weak coupling.  Another reason lies in the 
Z(N) reduced  effective actions~\cite{z3eff} and their plethora of undetermined parameters in the the case of gluodynamics: maybe their SUSY counterparts have more to say about his.

The thermodynamics of N=4 SUSY has been analysed about ten years ago~\cite{taylor}\cite{rey}\cite{tytgat}. These authors obtained weak coupling
results for the free energy and the Debye mass, in order to see how they connected to strong coupling results from the AdS/CFT approach. 

Recently other observables have been studied, like the mean colour-electric flux in the plasma measured by the spatial 't Hooft loop~\cite{armoni}. 

Their results for small coupling
 are one-loop precision. To this order nothing qualitative distinguishes the ${\cal N}=4$ case from gluodynamics. In this note we will compute the loop to the next order. This involves two-loop
calculations which are remarkably simple in gluodynamics~\cite{bhatta} and stay so in SUSY. However the results become in this order qualitatively 
different from gluodynamics.To two loop precision this effect is signaled by a divergence in the tension.
How to cure this divergence is discussed in section ~\ref{sec:discussion}.
This divergence is due to a gluino  zero mode~\cite{jackiwrebbi1976}.

When doing the calculation it is expedient to use the multidimensional formulation for SUSY theories~\cite{brink77}. Not only ${\cal N}=4$, but also ${\cal N} =2, 1$ cases are covered to the order we compute by a single parameter $\d$, the dimension used.

Our results may then be contrasted with those of strong coupling in ref.~\cite{armoni} in ${\cal N}=4$.

In section~\ref{sec:setup} the problem is set up and the results for the effective potential are stated. In section~\ref{sec:steep} some technicalities of our perturbative method to obtain the effective action are  given, in the next section~ \ref{sec:effaction} the physical reason for Casimir scaling is belaboured. 
Both of the latter  sections can be skipped by the reader who is imainly interested in the results.  The minimization of the effective potential to obtain the tension associated to the 't Hooft loop is done in section~\ref{sec:minimization}. The results are discussed in the next section and we wind up with conclusions. Some appendices are included to support claims in the main text.

 \section{The set-up and the results\label{sec:setup}}
 Consider the plasma of a four dimensional SU(N) gauge theory with only adjoint coloured matter, like gluodynamics or N=1,2 or 4 SUSY.  Such a theory 
 admits SU(N) gauge transformations  with a discontinuity in the center Z(N)
 of SU(N).  In particular  one can imagine a closed macroscopic spatial loop L  of Dirac magnetic flux~\cite{thooft78}. Such a loop is given by any static gauge transformation $V_k(L)$ that
 picks up a discontinuity $z_k=\exp(ik{2\pi\over N})$, whenever we circumnavigate the loop, say clockwise. In fig.\ref{fig:guises}(a) we show such a situation, with the loop in the (y,z) plane.  
 
 The claim is that the quantum average of such a loop measures the mean colour electric flux  through the loop. The quantum average is of course the
 Boltzmann trace, up to a normalizing factor:
 \be
\<V_k(L)\>= \Tr ~V_k(L)\exp(-H/T).
 \la{boltzmann}
 \ee
 
The trace is over all the colourless, i.e. gauge invariant states in our theory. This is the quantity we will compute 
to two loop order for SUSY theories. Like its homologue in gluodynamics~\cite{bhatta}  it has a rapid exponential fall-off, in fact with the area $A(L)$ of the loop:
\be
\langle V_k(L)\rangle=\exp(-\r_k A(L)).
\la{definitionr}
\ee

 The quantity $\r_k$ that controls this fall-off is called with a remarkable misnomer the "tension", or "dual string tension". 

Before we delve into formalism a few words on what we can expect on the basis of simple arguments.
The spatial t Hooft loop is the SU(N) version of the Dirac string in
quantum-electrodynamics.

 The Dirac string is engineered such that it is undetectable by quantum particles:  the wavefunction of any particle in the theory, boson or fermion,  is unaffected by the presence of the loop. The strength of the string is then equal
to the total magnetic flux coming out of the monopole at its end.  
If it has two endpoints it describes a monopole-antimonopole pair. The strength of the string is the number of times the phase 
of the gauge transformation winds around the gauge group U(1). For the 't Hooft loop the gauge group is $SU(N)/Z(N)$

The spatial 't Hooft loop  $V(L)$ is a macroscopic Dirac string $L$ closed on itself. From its unobservability  it would naively  seem to be an academic object. However this needs not to be true at low temperatures, where the macroscopic quantum-mechanical coherence of a condensate can create Abrikosov flux tubes. Under such circumstances~\footnote{Not occurring in the theories we will discuss here.} the spatial loop, or rather the state created by it from the ground state, will acquire an energy proportional to its perimeter.

There is a less well known role played by the spatial 
't Hooft loop. 
Mathematically, our closed Dirac string can rewritten through the use of Gauss' law as generator of gauge transformations
as an electric flux loop~\cite{stephanov99} (see also next subsection):
\be
V(L)=\exp(i4\pi{\int_{S(L)}d\vec S.\Tr \vec E Y_k/g}).
\la{electricfluxformula}
\ee
 $Y_k$ is a suitably chosen colour matrix in the fundamental representation
 , the hypercharge
  generating the Z(N) transformation ($\exp(i2\pi Y_k)=\exp(ik(2\pi/N)){\bf 1})$, and g is the gauge coupling. $\vec E$ is the electric field strength in the fundamental representation. 
  Since gluons are in the adjoint representation of $SU(N)/Z(N)$   their hypercharge is $\pm 1$ or $0$, whatever the value of k. What does depend on k are the multiplicities.
  
 Once more one would say off-hand that no quantum   
  particle in the theory
will detect the area. However one might argue that at very high temperatures the thermally screened particles are getting close to behaving as classical particles, escaping therefore the Dirac veto.
 This statement is correct in the limit of small coupling, where the  classical Debye screening length $ \l_D \sim {1\over{gT}}$ is very large compared to the de Broglie wave length $\l_B\sim {1\over T}$. 

%%%%%%%%%%%%%%%%%%%%%%%%%%
\begin{figure}[htb]
\begin{center}
\includegraphics[width=10cm]{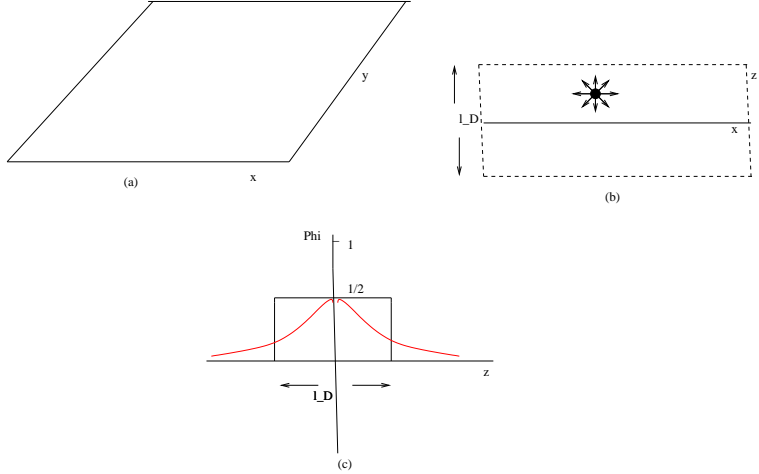}
\end{center}
%\centerline{\epsfig{file=plane.eps,width=10cm}}
\caption{Flux loop (a), Debye slab of thickness  $\l_D$ with particle emitting its flux (b), behaviour of flux as function of distance from the loop (red) in (c), and simplified (black)}
\la{fig:planecopy}
\end{figure}
%%%%%%%%%%%%%%%%%%%%%%%%%

    In that situation one can consider the slab (see fig. (\ref{fig:planecopy} (b)) to be a thin slab of width $\l_D$, where independent particles contribute to the flux through the loop. This is shown in fig. (\ref{fig:planecopy}(c).) It is then not hard to see~\cite{giovannaaltes01}  that one particle will contribute -1 to
 the flux (because only half of its flux shines through the loop, giving $\exp(i\pi)$), two independent  particles $(-1)^2$, etc..
    
    The Poisson distribution for a given species of screened gluons is that of Boltzmann particles in the fiducial volume -with on average  $\bar{l}=n\l_D\mbox{A(L)}$, n the density of the species-.
 It   reads:
 \be
P(l)={{\bar{l}}^l \over{l!}}\exp(-\bar l)   
 \ee  
 \noindent and so the flux average for this single species is:
 \be
  \sum_{l=0}^{\infty} P(l)(-)^l=\exp(-\l_Dn(T)\mbox{A(L)}).
 \ee
  
  Taking all species
  of gluons into account one has to determine how many 
  gluons in the adjoint are having a non-zero charge $Ad Y_k$ .   This turns out to be $2k(N-k)$. Because of the independence of the particles the exponent for one species is multiplied by this multiplicity:
 \be
 \r_k\mbox{A(L)}=2k(N-k)\l_Dn(T)\mbox{A(L)}.
 \la{classical}
 \ee
 $\r_k$ is the quantity controlling the area-off defined in eq. (\ref{definitionr}), called the tension.  It has nothing to do with a tension. The density $n$ is the same everywhere inside and outside the minimal area of the loop. It just
 tells us how the fall-off of the mean flux is controlled by the
  type of particles we have in the theory. The Casimir 
  scaling  is a signature for the adjoint representation. Evidently, in the confined phase we have no area law anymore as the colour electric flux is enclosed inside the hadrons.

 Coming back to gluodynamics, 
   in leading order in perturbation theory this result (\ref{classical}) was qualitatively reproduced~\cite{bhatta},  with the Casimir scaling factor $k(N-k)$ and 
 the factor $\l_D \sim 1/g$. And indeed on the lattice~\cite{defor} Casimir scaling is valid,  not only at very high  T, but    surprisingly almost down to the critical temperature! Of course the simple coupling constant behaviour in our leading  order  result (\ref{classical}) is only valid 
 at very high temperature. Lowering T will increase the coupling.  For gluodynamics the two loop results are known~\cite{bhatta}\cite{giovannaaltes01}, and so are the cubic terms~\cite{giovannaaltes02}. The cubic results are numerically small, but do violate Casimir scaling. Everything we know is consistent with the hypothesis that hard physics (on the scale T) respects Casimir scaling
and that soft physics (on the  scale gT) does not.

  For SUSY theories the gluons are accompanied by scalars
  and gluinos, all of them in the adjoint of SU(N). 
  So we  expect Casimir scaling according to the arguments above.

  In this paper we will push the perturbation series 
 to two loop order and will argue that the next, cubic, order is just a suitable rescaling of results known from gluodynamics.
 
 In next to leading order we will find again Casimir scaling. This
 is not surprising as we argue in section {\ref{sec:effaction}.
 But to three loop order one needs a dynamical mechanism to explain cancellations in the planar diagram sector.

 First we want to understand how the electric flux formula  (\ref{electricfluxformula})  comes about.

\subsection{What has the spatial loop to do with mean colour electric flux?}

First let us produce an explicit  version of such a singular gauge transform.

A set of Lie-algebra matrices $Y_k$ that generates the center through $z_k{\bf 1}=\exp(i2\pi Y_k)$  is~\cite{giovannaaltes01}:
\be
Y_k = {1 \over N} ~{\rm diag}(\underbrace{k,k,\dots,k}_{N-k~{\rm times}},
       \underbrace{k-N,k-N,\dots,k-N}_{k~{\rm times}})\nonumber. 
\la{defyk}
\ee
\noindent The k are integers.

The Gauss operator is written as:
\be
G^a=-E_i^bD^{ab}_i.
\ee
From the covariant derivative, $\vec D=\vec\pa+ig[\vec A,$, we have~\footnote{ Our notation is in terms of the SU(N) Lie algebra $A=A_a\lambda_a, \Tr\l_a\l_b={1\over 2}\d_{ab}, [\l_a\l_b]=if^{abc}\l_c$}
:
\ba
G^a&=&-\vec E^a.\vec\nabla+gf_{abc}A_i^bE^c_i \\
gj_0^a&=&-gf_{abc}A_i^bE^c_i .
\la{gaussnabelian}
\ea
The last equation defines the colour charge density $gj_0^a$.

Our singular gauge transformation is written as
\be
\exp(i\omega_a\lambda_a).
\la{numgaugetr}
\ee

Then we take  $\omega_a\l_a\equiv \alpha(\vec x) Y_k$ with $\alpha(\vec x)$  equal to half the solid angle defined between $\vec x$ and L. 

The 't Hooft loop is defined as the operator version of (\ref{numgaugetr}):
\be
V_k(L)=\exp(i \int d\vec x G^a\w_a).
\la{thooftloopdef}
\ee

We integrate inside a large sphere including the loop L the  operator density   
$\vec\nabla.(\vec E^a\omega^a)$ :

\be
\int dV\vec\nabla.(Tr\vec E\omega)= \mbox{surface term on $S_{\infty}$}-\int_{S(L)}d\vec S.Tr \vec EY_k 2\pi
\la{gaussna}
\ee
The surface term is negligeable because of the fall-off of the integrand.  In the second term the explicit form
of $\omega$ is used.

The l.h.s. is the sum of two terms:
\ba
\int dV\vec\nabla.(Tr\vec E\omega)&=&\int dV Tr(\vec\nabla.\vec E)\omega)+\int dV.(Tr(\vec E.\vec\nabla)\omega)\nonumber\\
&=& \int dVTr(\vec\nabla.\vec E -gj_0)\omega+\int dV Tr(\vec E.\vec\nabla +gj_0)\omega\nonumber
\la{gaussnaid}
\ea

We  have added and subtracted the colour charge density to
produce Gauss' operator (\ref{gaussnabelian}). So the first term cancels on physical states.
The second term equals, because of the normalization of the $\lambda$ matrices:
 \be
 \int dV Tr(\vec E.\vec\nabla +gj_0)\omega={1\over 2}\int dV (\vec E^a.\vec\nabla +gj^a_0)\omega^a).
 \ee
 It is the latter integral which is, up to a sign,  the gauge transformation     
 (\ref{gaussnabelian}) and hence,
combining  eq. (\ref{gaussna}) and (\ref{gaussnaid}) and exponentiating we get:
\be
\exp({4\pi\over g} i\int_{S(L)}d\vec S.Tr\vec EY_k)=\exp(i{1\over g}\int dV (\vec E^a.\vec\nabla +gj^a_0)\omega^a)).
\ee
\noindent  in the physical Hilbert space.
The r.h.s. is the gauge transformation in Hilbert space of $\exp(i\omega(\vec x))$ with a discontinuity $z_k$ in the centergroup on the surface. This was the definition of the 't Hooft loop. On the left hand side we see its representation as an electric flux operator.

Once its role as electric colour flux operator is understood, it is intuitively clear that the average (\ref{boltzmann}) will be  an area law in a plasma~\cite{giovannaaltes01}:
\be
\<V_k(L)\>=\exp(-\r_k(T)\mbox{Area(L)}).
\la{area}
\ee
The quantity that controls the fall-off is called the electric tension, or "dual tension". 

Now we  return to its quantum average (\ref{boltzmann}) and discuss two other 
guises for this average.

\subsection{Two other guises for the spatial 't Hooft
 loop average}
 
 The second guise is useful for calculations on the lattice or in perturbation theory and is the Euclidean path integral version.  The gauge transformation is replaced  by the Polyakov loop in the Euclidean time direction $\t$:
\be
 P(A_0)(\vec x)={\cal{P}}\big( \exp(i\int_0^{1/T}d\t A_0(\vec x, \t))\big).
\la{defpol}
 \ee
 
 Since it transforms under a gauge transformation as an adjoint:
 \be
  P(A_0^{\W})=\W(\vec x,0)P(A_0)\W^{\dagger}(\vec x,1/T),
 \la{poltransf}
  \ee
  \noindent the N-1 phases of the  diagonalized loop are invariant under periodic
  transforms, and shift under periodic modulo centergroup transformation~\cite{pis1981}.
  
  So the gauge invariant content of the loop can be stored in the trace of N-1 powers of the loop:
  \be
  P^{(k)}(A_0)=\Tr P(A_0)^k,~~\mbox{k=1,......,N-1}.
  \la{invts}
 \ee

Imagine the Polyakov loop along a straight  line in the x-direction. The straight line pierces the area inside the loop L. in the (y,z) plane. As long as we stay away from the boundary the  Polyakov loop will have the same profile on any such line, and will only depend on z.  
When the Polyakov loop crosses the surface, where the loop operator $V_k(L)$
has its discontinuity, it will pick up the discontinuity because of (\ref{poltransf}).

So what we need is the probability for a given Polyakov loop profile C(z)  to appear.  C(z) is a diagonal traceless real NxN matrix, obeying
\be
P^{(k)}(C)=\Tr\exp(ikC/T).
\la{profileC}
\ee

  Then
we maximize this probability with the boundary conditions at  infinity $P(A_0)(x=\pm\infty)={\bf 1}$ and by the jump at the surface.

The probability is determined by the Euclidean path integral with  the constraint:

 \be
{\cal C}= \Pi_{k,x}\d\big(\overline{P^{(k)}(A_0)}-\Tr\exp(ikC/T)\big).
 \la{constraint}
\ee
 The overlined Polyakov loop is a short hand for the average over
 y and z coordinates. For this to make sense we put the system in a 
 finite but large transverse volume $V^{tr}$ of the y and z coordinates.
 
 The constraint is inserted in the path integral. We then get, up to the 
usual partition function for the probability that a given profile in the x direction develops:
\be
\int DADSDF ~~{\cal C}\exp(-S)=\exp(-V^{tr}U(C)-\mbox{bulk free energy}).
\la{constrainedint}
\ee
We integrate over gauge fields A, scalar fields S and fermionic fields F.

$U(C)$ is the resulting  effective action. %As the  constraint does not respect SUSY there is an extra source of SUSY breaking in the domain wall.
The effective action can be written in terms of a kinetic term $K$  proportional to $\Tr(\partial_zC)^2$ and an effective potential $V(C)$:
\be
U(C)=\int dx\Big({1\over{2g^2N}}\widehat K\Tr(\partial_zC)^2+V(C)\Big).
\la{effactkv}
\ee
The (y,z)  plane at x=0, where the Polyakov loop jumps, is called a domain wall. This is a little confusing, because we think of a domain wall as a soliton, living sub  specie aeternitatis. With only a reinterpretation
 of the Euclidean coordinate axes we can obtain such a picture, at the price 
 of having either y or z axis as the infinite time axis.
 %%%%%%%%%%%%%%%%%%%%%%%%%%%%%%
  \begin{figure}
\begin{center}
\psfig{file=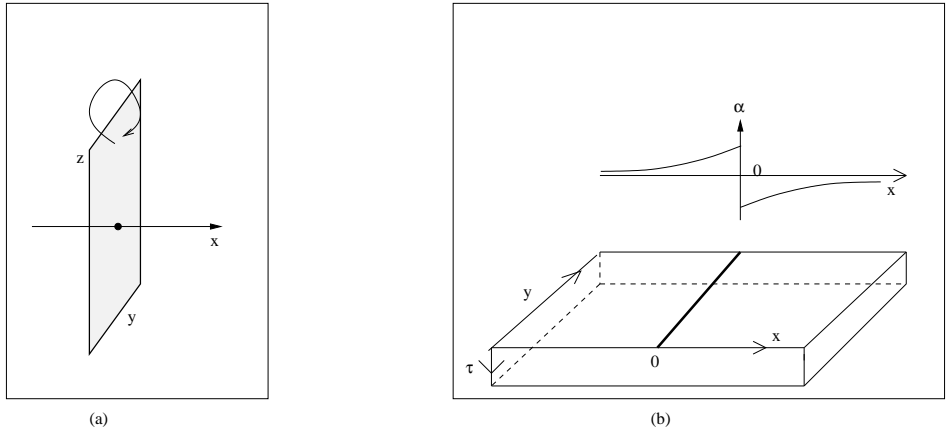,angle=0,width=12.5cm}
\end{center}
\caption{Left panel: the spatial 't Hooft loop in the (y,z) plane at
x=0. The arrow on the upper part indicates the clockwise 
circumnavigation defining the jump, eq. (\ref{thooftloopdef}). On the right the same loop but now with the z-axis interpreted as time axis. The (x,y) plane contains a one-dimensional wall at x=0, and a microscopically small, periodic spatial $\t$-direction along which the Polyakov loop profile $P(C)=\exp(i 2\pi T\a(x)Y_k)$ is winding, with the  phase $\a$ of the profile as function of x shown in panel (b).
Fermionic fields are supposed to be anti-periodic in the $\t$ direction.}
\la{fig:guises}
\end{figure}
 %%%%%%%%%%%%%%%%%%%%%%%%%%%%%%
 
 Hence the third guise is that of a one dimensional defect, 
 that lives in the two dimensional (x,y) world with a small periodic space dimension labeled $\t$ (see fig. (\ref{fig:guises})(b)) . The Euclidean time is now the z-axis, so the temperature is zero.  The defect is given by the jump the Polyakov loop  looping the small 
 periodic space dimension $\t$  feels when crossing the line defined by x=0 in the (x,y) plane.   Thus the defect is once more a magnetic flux.% but now of the type$F_{\t,x}$. 
 The energy per unit length of the magnetic flux is again $\r_k(T)$. It is created at say $z=-L_z/2$ and lives on  in the z direction during the time $L_z$ , that we let go to infinity~\cite{thirdguise}.  Calculationally there is no difference with the second guise, the Euclidean path integral (\ref{constrainedint}).

  %The Hamiltonian $H(A_{0,x,y},\Pi_{0,x,y})$ with z as time
% together with the  operator $V_k(L_y)$   %${\cal C}$ 
% give the same Euclidean constrained path integral  as before. 
% creating through a singular gauge transformation in the $\t,x$ plane the magnetic flux  has the same zero temperature average:
% \ba
%&& \tr V_k(L_y)\exp(-L_zH(A_{0,x,y},\Pi_{0,x,y})~~\mbox{and}\nonumber\\
% &&\Tr V_k(L)\exp(-H/T)\nonumber
% \ea
% \noindent 
% The two traces give the same $\r_k(T)$.  The one with the Euclidean time z
% creates a an elongated loop in the z direction $L_z>>L_y$~\cite{thirdguise}. 

Its usefulness lies amongst more in  an alternative but simpler and elegant  understanding of the nature of the divergencies we will meet  in the kinetic part of the effective action.

\subsection{Results for the effective action for ${\cal N}=1$, 2 and 4 SUSY\la{sec:resultseffaction}} 

For our purpose we follow Vasquez-Mozo~\cite{vasquez} and compute with the ${\cal N}=1$ SUSY action formulated in $\d\ge 4$ dimensions~\cite{brink77}. When $\d=10$ the fermions are constrained to be
Majorana-Weyl fermions, and we recover N=4 SUSY. When 
$\d=6$ the fermions are Weyl fermions and we recover ${\cal N}=2$ SUSY. When $\d=4$ we have by definition ${\cal N}$=1 SUSY, with Majorana fermions.
So generically the action we use reads:
\be
S=\int d^4x\bigg({1\over 2}\Tr F_{pq}^2+\Tr\bar\l\G_pD_p\l\bigg)~~p, q=0,1,...,\d -1..
\la{n1susyaction}
\ee
The fields in our action are the bosonic ones $A_p$, p=0,1...9, appearing in the field strength $F_{pq}=\pa_pA_q-\pa_qA_p+i[A_p,A_q]$. Both 
fermions and bosons are  matrix-valued (in the Lie-algebra of SU(N), and the covariant derivative reads $D_p=\pa_p+i[A_p,$. 

The fermions $\l$ are spinors with $\d-2$
independent components because of the  constraints. Their 
dimension (and that of the Dirac matrices $\G_p, p=0,1,...\d-1$) is $2^{\d/2}$. All fields only vary with $x_0\equiv\t,x_1=x,x_2=y,~\mbox{and}~ x_3=z$, the remaining dimensions serve
as bookkeeping device for the flavours (SU(4) for $\d=10$, and $U(2)$ for $\d=6$).  In bosonic computations the $\d$-dimensional unit matrix ${\bf 1}_b$
often appears, whereas in their fermionic counterparts the projector matrix
${\bf 1}_f\equiv \l\times\bar\l$ ($2^{{\d\over 2}}$ dimensional) with $\Tr{\bf 1}_f=\d-2$ is present. For details see Appendic C.

\subsubsection{Some basic facts and notation}
We use throughout dimensional regularization with $d=4-2\e$, $\e\rightarrow 0$.

The Feynman rules are the Euclidean finite temperature 
rules, with the Matsubara frequency $l_0=2\pi n T$, n integer for bosons, and
$l_0+\pi T$ for fermions. The constraint (\ref{constraint}) will
introduce a background $A_0=C+gQ_0$. Therefore the covariant derivative $D_0=\pa_0+ig[C,$ will remain in the 
in the propagators. Hence it is  advantageous to go into the Cartan basis for all the quantum fluctuations $gQ_p=A_p$, for $p\ge 1$ and  for the quantum fluctuations 
$\l$. The Cartan basis consists of the off-diagonal matrices $\l^{ij}$ with the (ij) entry equal to $1/\sqrt{2}$ and other entries zero. They form an eigenbasis for C:
\be
[C,\l^{ij}]=(C_{ii}-C_{jj})\l^{ij}.
\la{cartan}
\ee  

The potential is known, on general grounds~\cite{gocksch}, to have absolute minima at the locations where $\exp(iC/T)$
takes value in the centergroup.

More specifically~\cite{giovannaaltes01}  the potential valleys  of the lowest 
order (one-loop) potential $V_1$ relating the origin to the nearest minimum with Z(N) value $z_k$ are described by the matrices (see (\ref{defyk}):
\be
C=2\pi T q (x)Y_k.
\la{valley}
\ee
The scalar function $q(x)$ is to be determined as function of x from the minimization of U(C). It enters the Matsubara frequencies~\footnote{Not only in the propagators but also in the vertices!} as: 
\ba
&~&l_0+2\pi Tq~~\mbox{ for bosons and}\\
&~&l_0+\pi T +2\pi Tq~~\mbox{for fermions.}
\la{qshift}
\ea

Thus, inside the wall the role of  bosons and fermions starts to get interchanged. In particular in the center of the wall-where $q=1/2$- the fermions
develop long range behaviour in the two dimensions left,  just as the bosons do outside the wall at q=0 and 1, but now in three dimensions. We will come back to this phenomenon in section \ref{sec:discussion}.

The summation/integration measure is written as
\be
\sumint_{~l(b,f)}\equiv T\sum_n\int {d^{d-1}l\over{(2\pi)^{d-1}\mu^{2\e}}},
\ee
\noindent  where the suffix b or f refers to whether we put 
the bosonic or fermionic Matsubara frequency. The scale 
$\mu$ starts to play a role in the one loop kinetic term.

We fix the notation for the sum/integrals we meet in the one and two loop. They are proportional to Bernoulli polynomials
and read ($l_q^2\equiv (l_0+2\pi Tq)^2+\vec l^2$):
\ba
\sumint_{~l(b)} \log(l_q^2)&=&\hat B_4(q)\\
\sumint_{~l(b)}  {(l_0+2\pi T q)\over {l_q^2}}&=&\hat B_3(q)\\
\sumint_{~l(b)} {1\over{l_q^2}}&=&\hat B_2(q)\\
\sumint_{~l(b)} {(l_0+2\pi T q)\over{l_q^4}}&=&\hat B_1(q).
\la{bernoullihat}
\ea
The right hand sides are represented by Bernoulli polynomials $B_r$ ( defined in Appendix B)  on the interval $-1\le q\le 1$.
If the trace is fermionic, the argument of the B-polynomials is shifted by ${1\over 2}$. The integrals on the left hand side
are periodic mod 1 in the argument q. %Their representation in terms of Bernoulli polynomials is only valid on the interval $0\le q\le 1$. Even (odd) polynomials are even (odd) under the interchange  $q\ra 1-q$. 

Obviously one recovers by 
differentiation  $\hat B_3$ from $\hat B_4$,  etc., up to some overall factor. As they are periodic "polynomials" they are
not analytic at $q=0, 1....$.  

Since the breaking of supersymmetry resides in the shift from bosonic to fermionic traces, the non-renormalization of supersymmetric potentials makes
differences like
\be
D_r(q)=\hat B_r(q)-\hat B_r(q+{1\over 2})
\la{difference}
\ee
\noindent occur frequently. 
The number of bosonic (or fermionic) degrees of freedom
appears naturally as well. We write
\be
\D={\d-2\over 2},
\la{dof}
\ee
\noindent equal to 1, 3 and 4 for respectively ${\cal{N}}=1,2,~\mbox{and}~ 4$.

\subsubsection{Results for the effective potential}
From refs~\cite{taylor}\cite{vasquez}\cite{tytgat} we quote 
the one and two loop free energy:
\ba
F_1&=&\D(N^2-1)\hat D_4(0)\\
F_2&=&\left(\D\tilde g\right)^2(N^2-1)\hat D_2(0)^2,
\la{freeenergy}
\ea
\noindent with $\tilde g^2$ the 't Hooft coupling.

The dimension of the adjoint representation appears because each member of a given multiplet has the same weight in the free energy.

The screening masses of the gauge fields are from the same references:
\be
 m_D^2={\D\over 2}\tilde g^2T^2.
 \la{screeningmassgauge}
 \ee
%%%%%%%%%%%%%%%%%%%%%%%%%%%%%%
\begin{figure}
\begin{center}
\psfig{file=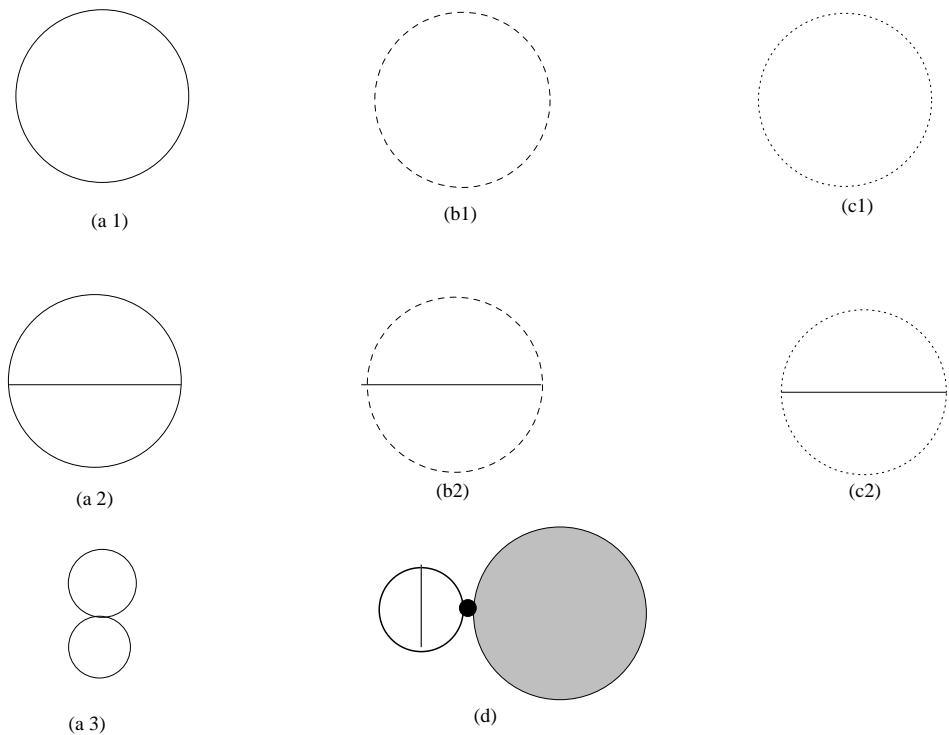,angle=0,width=12.5cm}
\end{center}
\caption{One and two loop contributions to the effective potential. Continuous lines are bosons, broken lines are fermions and dotted lines are ghosts. In (a) the bosonic contributions, in (b) the fermionic contributions and (c) the ghost contribution are shown. In (d) 
 the Polyakov loop (fat circle), renormalized by one gluon exchange, is inserted into the sum of the loops (a1), (b1) and (c), given by the shaded blob.}
\la{fig:susyonetwolooppot}
\end{figure}
%%%%%%%%%%%%%%%%%%%%%%%%%%%%%% 
The potential terms of the effective action read on the interval $0\le q\le {1\over 2}$:
 \ba
 V&=&V_1+\tilde g^2V_2\nonumber\\
 V_1&=&2k(N-k)\D\hat D_4(q)\\
 V_2&=&k(N-k)\D\bigg[\D\bigg\{\bigg(\hat D_2(q)+\hat D_2(0)\bigg)^2-\hat D_2(0)^2\bigg\}+4\hat B_1(q)\hat D_3(q)\bigg]
 \la{effpot}
 \ea
 
 The contributions of the individual diagrams are given in Appendix D.

Note the factors $k(N-k)$. They replace the factor $N^2-1$
in the free energy and are referred to as "Casimir scaling". 
To one loop order they appear as a consequence of simple counting: for a given charge $Y_k$ there are k(N-k) excitations in the adjoint with charge 1 and an equal number with charge -1. As the $\hat B_4(\pm q)$ take the same value we have Casimir scaling.

For the two loop potential this counting is not enough as there are interactions. In section \ref{sec:effaction} the  double line notation is used to show how Casimir scaling is nethertheless realized.

 The factors $\D={\d-2\over 2}$ count the number of $\d$ dimensional loops. The last term of $V_2$ is the exception, for a simple reason. This latter term is due to the one loop renormalization of the Polyakov loop (see fig.(\ref{fig:susyonetwolooppot}~(d)), which {\it only} involves the gauge sector and is proportional to $B_1$:
 \be
 \d q=-{\tilde g^2\over {(4\pi)^2}} (3-\xi)B_1(q)
 \ee
  Inserted into the 
result for  $V_1(q)$, $V_1(q+\d q)$, it gives a result linear in ${\d-2\over 2}$ and $\hat B_3$. We show in more detail in section \ref{sec:steep}
how these insertions come about.
 
 To obtain the old result for gluodynamics, put $\d=4$ and drop the fermionic contributions to find~\cite{bhatta}:
 \be
 V_1+\tilde g^2V_2=k(N-k){4\pi^2\over 3}T^4q^2(1-q)^2\bigg[1-5\bigg({\tilde g\over{4\pi}}\bigg)^2\bigg].
\la{gluodynpot}
\ee

%%%%%%%%%%%%%%%%%%%%%%%%%%%%%%%%%%%%%
\begin{figure}
\begin{center}
\psfig{file=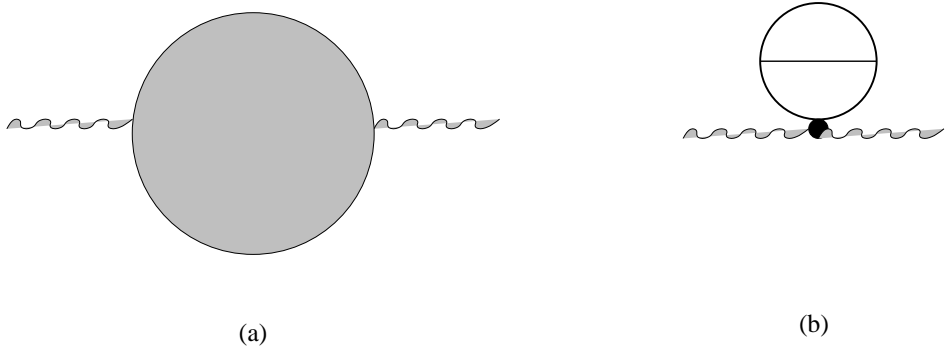,angle=0,width=12.5cm}
\end{center}
\caption{The contributions to the one loop effective kinetic term. The wiggly lines are $q$  lines. The  shaded blob in (a) is the sum of one loop boson, fermion and ghost graphs. In panel (b) the renormalization of the Polyakov loop is shown.}
\la{fig:susyoneloopkin}
\end{figure}
 %%%%%%%%%%%%%%%%%%%%%%%%%%%%%%%%%%%%%%

\subsubsection{One loop kinetic term}

We now turn to  graph (a) in fig. \ref{fig:susyoneloopkin}. 
It is obtained by computing the determinant of boson  ghost fields in
(\ref{quadratic}) and expanding the determinant in terms of derivatives of the background. 
We suppose the colour degrees of freedom have been diagonalized in the Cartan basis $Q^{ij}$, so that the background $B$ below stands for $B^{ij}=B_{ii}-B_{jj}$.

So we have to take into account that the x-dependence in the background does not commute with the momentum operator $P_x={1\over i}\pa_x$ in the d'Alembertian. The $B$ field will depend on the operator $X$, $\underline{B}\equiv B(X)$, that  obeys:
\ba
[P_x,X]&=&-i~~\mbox{and}\nonumber\\
X|x\>&=&x|x\>.
\la{pxrelations}
\ea

We do this in Feynman gauge $\xi=1$. This gives for the bosonic part of the effective kinetic term ( for the fluctuation matrix see (\ref{quadratic}):
\be
K_b=\Tr_b{1\over 2} \log\bigg(- D^2(\underline{B}{\bf 1_b}-2F_{0x}{\bf 1_{0x}}\bigg)-\Tr_b\log\bigg(-D^2(\underline{B})\bigg) .
\la{trlogx}
\ee

The first term is due to vector bosons and scalars, the second logarithm
due to the ghost.
The magnetic moment term of the gluonic component  is proportional to $F_{0x}=B'$ and hence $\OO(g)$ and expanded out.  The scalar component has no such term, as reflected in the matrix:
\be
{\bf 1_{0x}}\equiv\mbox{Diag}(1,-1,0,...,0).
\ee
 So $K_b$ becomes:
 \be
 K_b={\d-2\over 2}\Tr_b \log(- D^2(\underline{B}))-{1\over{2!}}F_{0x}^2 4Tr_b{1\over{(-D^2(B))^2}}\bigg).
\la{trlogxb}
\ee

We do the same for the fermionic determinant. The result  there is  different
in that the spin terms are present for {\it all} fermionic components~\footnote{ Easily checked by going into an explicit representation for the $G_0$ and $G_x$ matrices. See also Appendix C.}.

 One gets accordingly:
\ba
K_f&=&-\Tr_f{1\over 2}\log\bigg(- D^2(\underline{B}){\bf 1_f}-F_{0x}\G_{0x}{\bf 1}_f)\bigg)\nonumber\\
&=& -{\d-2\over 2}\bigg(Tr_f\log(-D(\underline B)^2-{1\over{2!}}F_{0x}^2Tr_f{1\over{(-D(B)^2)}}\bigg).
\la{trlogxf}
\ea
The spin matrix $\G_{0x}={1\over{4i}}[\G_0,\G_x]$.

The trace is mixed in x and conjugate momentum, instead of the pure momentum trace  $\sumint$  in section \ref{sec:setup}:
\be
\Tr O\equiv=T\sum_n\int{d^{d-2}l\over{(2\pi)^{d-2}\mu^{2\e}}}\int dx\<x|O|x\>.
\ee 
 The suffixes b or f refer to treating the Matsubara frequency $l_0=2\pi nT$ as bosonic or fermionic, by adding in the latter case $\pi T$.
 
The total effective kinetic term is now:
\be
K=K_b+K_f 
 \la{total}
 \ee
 \noindent and from (\ref{trlogxb}) and  (\ref{trlogxf}) follows:
 
 \ba
 K&=&{\d -2\over 2}\bigg(\Tr_b\log (-D^2(\underline{B}))-\Tr_f\ log(- D^2(\underline{B}))\bigg)\nonumber \\
 &+&{1\over {2!}} F_{0x}^2\bigg(-4 Tr_b{1\over({-D^2(B))^2}}+{\d-2\over 2}Tr_f{1\over{(-D^2(B))^2}}\bigg)
 \la{total2}
 \ea

 The short distance behaviour of the total effective kinetic term is independent of the background and  whether the trace is bosonic or fermionic .  The short distance contributions of the first two terms in $K_b$ and $K_f$~\footnote{ Working out their  contributions proportional to $F_{0x}^2={B^{\prime}}^2$ is straightforward and is done in Appendix A}  do therefore cancel in the   effective kinetic term, and we conclude the short distance behaviour  of the kinetic part of the SUSY action is fixed entirely by the difference of the magnetic moment terms of bosons ($-4$)
  and fermions (${\d-2\over2}$):
\be
K_{shortdistance}\sim{1\over {2!}}F_{0x}^2((-4+{\d-2\over 2})({1\over{\e}}+ \OO(\e)).
\ee

This difference vanishes for $\d=10$, as it should. If $\d$ is smaller than 10 we retrieve the asymptotically free SUSY theories. For $\d$ larger than 10 asymptotic freedom is lost.

Finally, from (\ref{resultapp}) in Appendix A and (\ref{total2}) the kinetic term
equals in $d\rightarrow 4$  and $\d$  dimensions:
\ba
K&=&k(N-k){(2\pi Tq'(x))^2\over{\tilde g^2}}\times\nonumber\\ &\times&\bigg [1+
{\tilde g^2\over{(4\pi)^2}}{\G({5-d\over 2})\over{\pi^{{5-d\over 2}}}}\bigg({T\over {\mu}}\bigg)^{-2\e}\bigg\{%b_0(\d)~{1\over \e}+
\bigg(b_b(\d)+{2\e\over 3}{\d-2\over 2}\bigg)\sum_n{1\over{|n+q|^{5-d}}}\nonumber\\
&-&\bigg(-b_f(\d)+{2\e\over 3}{\d-2\over 2}\bigg)\sum_n{1\over{|n+1/2+q|^{5-d}}}\bigg\}-4{\tilde g^2\over {(4\pi)^2}} \bigg]
\la{kinetic}
\ea
 The very last $\OO(\tilde g^2)$ term comes from diagram (b) in fig.(\ref{fig:susyoneloopkin}) and results from substituting the renormalization $\d q=-{\tilde g^2\over {(4\pi)^2}} (3-\xi)B_1(q)$ ( see eq. (\ref{loopvertex})) into the kinetic term $q'^2$ in Feynman gauge $\xi=1$~\footnote{Including this renormalization the $\xi$ dependence drops out from the kinetic term as discussed in ref.~\cite{bhatta} and \cite{korthals94}}. Details can be found in section \ref{sec:steep}.
The one loop $\b$-function coefficients $b_b$ respectively  $b_f$ are 
the contributions of the bosons  respectively fermions:
\ba
b_b(\d)&=&-{11\over 3}+{\d-4\over 6}\\
b_f(\d)&=&{\d-2\over 3} 
 \la{bfunction}
 \ea

The total $\b$-function is then $b_0(\d)=b_b(\d)+b_f(\d)={\d\over 2}-5$, asymptotically free for $\d < 10$.

Substitute~\cite{rg}:
\be
\sum_n{1\over{|n+q|^{5-d}}}={1\over {\e}}-(\psi(q)+\psi(1-q))
\ee
\noindent in (\ref{kinetic}).

In terms of the %$\overline{MS}$
renormalized coupling  $g(T)$
\be
{1\over{\tilde g^2(T)}}={1\over{\tilde g^2}}\bigg\{1+\bigg({\tilde g\over {4\pi}}\bigg)^2b_0(\d)\bigg[{1\over \e}-\log{\pi}-2\log\bigg({T\over\m}\bigg)\bigg]\bigg\}
\ee
 and $\d$, we obtain for the kinetic  term:%in (\ref{effactkv}):
\ba
K&=&k(N-k){(2\pi Tq'(x))^2\over{\tilde g^2}(T)}\bigg(1-\bigg({\tilde g(T)\over{4\pi}}\bigg)^2\bigg\{b_b(\d)(\psi(q)+\psi(1-q))+\nonumber \\
&+&b_f(\d)(\psi(q+{1\over 2})+\psi(-q+{1\over 2}))-b_0(\d)\psi(1/2)-4\bigg\}\bigg).
 \la{renkinterm}
 \ea
 
  The representation in terms of the logarithmic derivative of the $\G$ function  $\psi$ (see \cite{rg}) is valid for $0\le q\le {1\over 2}$. The reader will find the details in section \ref
{sec:effaction} and Appendix A.

 \section{Steepest descent of the constrained path integral\label{sec:steep}}
 
 This section will  deal with the constraint (\ref{constraint}) in the Euclidean path integral. The main point is the systematic approach,
 which is of import  for higher loop calculations. For readers who are only interested in our two loop results this section can be skipped. 
 
 We reproduce for discussion's sake (\ref{constrainedint}): 
 \be
\int DADSDF ~~{\cal C}\exp(-S)=\exp(-V^{tr}U(C)-\mbox{bulk free energy})
\la{constrainedint1}
\ee
\noindent and the constraint ${\cal C}$:
 \be
{\cal C}= \Pi_{l,x}\d\big(\overline{P^{(l)}(A_0)}-\Tr\exp(ilC/T)\big).
 \la{constraint1}
\ee

The steepest descent expansion of the Euclidean path integral is straightforward, and in principle there is nothing new except the handling of the constraint ${\cal C}$, in (\ref{constraint1}). This new feature is discussed in this section, in particular the diagrams in figs.(\ref{fig:susyonetwolooppot} (d)) and 
(\ref{fig:susyoneloopkin} (b)) in the second subsection.

\subsection{Generalities on expanding the constraint}
    The most obvious way of dealing with the constraints ( which are only depending on the space variable x) is to Fourier
    analyze them with one-dimensional x dependent fields $\g^{(l)}(x)$ where l runs through the integers from 1 to N-1. Since the constraints are gauge invariant, so are the constraint field $\g$. We then get an action $S_{con}$ with these constraint fields included, 
    
\be
S_{con}=i\sum_l\int dx \g^{(l)}(x)\bigg(\overline{P^{(l)}(A_0)}(x)-\exp(ikC(x)\bigg)+ S. 
\la{scon}
\ee
The constrained action appears in units of the squared coupling $g^2$ in the path integral:
\be
\int DA DS DF D\g\exp(-{S_{con}\over{g^2}}).
\la{constrainedintg}
\ee
We start the steepest descent around the field C(x) by introducing background fields $B(x)$ and $\g_c^{(l)}(x)$ in the potential and the 
constraint fields:
\ba
A_0&=&B(x)+gQ_0,\\
\g^{(l)}(x)&=&\g_c^{(l)}(x)+g\g^{(l)}_q(x).
\la{backgrounds}
\ea
$B(x)$ is diagonal, traceless of size $N\times N$. The subscript q
refers to the quantum fields that fluctuate around the gauge invariant sources  $\g_c^{(l)}(x)$ and 
$Q_0$ fluctuates in four dimensions. around the background field $B$. The lattter is not gauge invariant.

The expansion in the quantum fields in $S_{con}$  yields equations of motion that fix
$\g^{(l)}_c$  and $B$ in terms of $C$, and define a minimum of $S_{con}$. They read respectively:
\ba
\bigg(P^{(l)}(B)-\exp(ilC/T)\bigg)&=&0\nonumber\\
 i\g_c^{(l)}lP^{(l)}_{,d}(B)+{\partial_x}^2B_d&=&0.
 \la{linear}
 \ea
 Derivatives with $\overline{Q^0_d}(x)$ are indicated with subscripts
 $,d$. Only diagonal and  overlined Q's are relevant, because B is diagonal and both B and $\g^{(l)}$ do only depend on x. The matrix 
 $t^{(l)}_d(C)\equiv lP^{(l)}_{,d}(B)$ brings us from the d-basis,  numbering the diagonal elements of the Lie-algebra, to the l-basis, numbering the powers of the unitary Polyakov loop. It reflects the non-linearity of the constraint. Surprisingly, in amplitudes it will cancel, as we will see in the next subsection.

 So B=C from the first equation, and the second equation says the d component of $i\g^{(l)}_c$ equals the d-component of  the double gradient of B (or C). From the first linear equation follows that only
 \be
 {1\over{g^2}}\Tr ({\partial_xC})^2
 \ee
 remains as classical term. So there is no classical potential term. This has an important consequence:
 the potential comes entirely from quantum corrections, which are 
  of $\OO(g^0)$. As  a consequence the gradient of the C field must  be soft:
  \be
  \partial_xC(x)\sim g.
  \la{soft} 
  \ee

  So the second linear equation in (\ref{linear}) tells us that the gauge-invariant source is soft as well; the $\g_c^{(l)}$ are      $\OO(g^2)$.

 Turning to the quadratic terms in $S_{con}$ we have in  linear $\xi$ gauge the following terms:
 \ba
 &~&Q_{\mu}\bigg( -D^2(C)\delta_{\mu\nu}+(1-{1\over{\x}})D_{\mu}D_{\nu}-2i[F_{\mu\nu}\bigg)Q_\nu  +\mbox{ghost terms}\\
 &+&i\g^{lk)}_qkP^{(l)}_{,d}(C)Q^0_d\\
 &+&i\g_c^{(l)}P^{(l)}(C)_{,ab}Q^0_aQ^0_b
 \la{quadratic}
 \ea
 
 We write in the sequel $Q^{(l)}=kP^{(l)}_{,d}(C)Q^0_d$.

We have kept  in these terms {\it all}	 $C$ dependent terms, although their derivatives  are soft, like the last term with the gauge invariant source.
 
The first term is the familiar inverse background propagator, the second one, after integration over the constraint fields, gives us delta function constraints on the $Q^{(l)}$. 

The third one, because of the softness of the source $\g_c^{(k)}$,  will give a one loop
correction to the classical term $(\partial_zC)^2$ as we will see in the next  subsection and is entirely due to the constraint. It introduces a renormalization of the background field in the kinetic term.  %This renormalization  is the only one at this order  that does not disappear when we impose SUSY boundary conditions. It refllects the SUSY breaking effects of the constraint.

The new feature entering in cubic and higher order terms is the appearance of terms proportional to
$i\g_q^{(l)}$, e.g. $ i\g_q^{(l)}P^{(l)}_{,ab}Q_aQ_b$.

 If we want to integrate over the fluctuations $\g_q^{(l)}$ to reinstaure the constraint on the quantum fields $Q_0^{(l)}$  we can replace the factors;
\be
i\g_q^{(l)}\exp(i\g^{(l)}_qQ^{(l)})\rightarrow {\partial\over{\partial Q_0^{(l)}}}\exp(i\g^{(l)}_qQ^{(l)})
\la{insert}
\ee
Integrating now over the $\g_q$   we get the constraints on the $Q_0$. The partial derivative $ {\partial\over{\partial Q_0^{(l)}}}$ acts through partial integration on the vertices of the usual Yang-Mills action
with gauge fixing and ghosts and creates new, "inserted" vertices. Such vertices are shown as fat dots in fig.(\ref{fig:susyonetwolooppot} (d)).

\subsection{Insertions to two loop order}

Let us now look at what these insertions of the renormalization of the Polyakov loop imply for the two loop calculation.

We start this  with a useful identity~\cite{korthals94} for this renormalization, and the two corrections to order $g^2$ that follow immediately from this identity.  

The Polyakov loop is shown in fig. (\ref{fig:wilsononeloopaverage}) with one gluon going across.
This is the lowest order $g^2$. On the righthand side are shown the 
ways the gluon can go across. Ignore the partial derivative
with $Q^{(l)}_0$ for the moment. If one adds all the possible exchanges
the result is :
\ba
\langle P^{(l)}_{,ab}(C)Q^0_aQ^0_b\rangle&=&t^{(l)}(C)_{,d}\delta_2C_d, ~\mbox{with}:\\
\delta_2C_{ii}-\d_2C_{jj} &=-&{\tilde g^2\over{(4\pi)^2}}B_1(C_{ij})(3-\xi).
\la{loopvertex}
\ea

The first identity  causes the simplicity  of the one-loop insertions, because all the non-linearities of the constraint are sitting in 
the transition matrix $t^{(l)}_d(C)$.  In fact, as promised, it will disappear from the insertion graphs in fig. (\ref{fig:susyonetwolooppot})
and (\ref{fig:susyoneloopkin}), once we look at the graph as a whole. 

For the insertion graph in the potential term the partial derivative acts on one of the gauge field vertices. But since we sum over l the transition  matrix in $Q^{(l)}_0$ cancels with
the one in (\ref{loopvertex}) and we are left with $\d_2C_d{\pa\over {\pa Q^d_0}}$ for the insertion graph.

To find the  insertion graph for the kinetic term
 apply our identity to the third term in (\ref{quadratic}) and combine the result with the second equation in (\ref{linear}):
 \ba
&~& i\g_c^{(l)}\langle P^{(l)}(C)_{,ab}Q^0_aQ^0_b\rangle=-(3-\xi)\partial_x^2C_d.\delta_2C_d=-(3-\xi)\pa_xC_d\pa_x f^{d,ij,ji}\hat B_1(C_{ij})\\
&=&(3-\xi)\Tr[\pa_xB,[\l_{ij}\l_{ji}]]\pa_x\hat B_1(C_{ij})\\
&=&(3-\xi){1\over 2}\pa_xC_{ij}\pa_x\hat B_1(C_{ij})
\la{insertkinetic}
\ea

Now use that $\widehat B_1$ ( given in  appendix B) is linear in its argument :
\be
\hat B_1(B_{ij})=-{T\over{4\pi}}(B_{ij}-{1\over 2}\e(B_{ij}).
\ee

The derivative of the first term is what we retain in the graph.
The derivative of the second term equals $\delta(C{ij})\pa_x B_{ij}$. It can be ignored in the tension because of the equations 
of motion for $B$ (see section \ref{sec:minimization}).

\begin{figure}
\begin{center}
\psfig{file=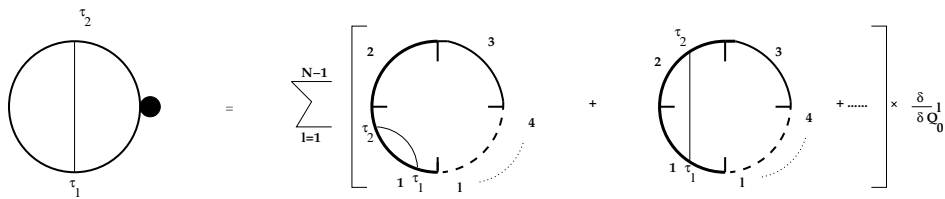,angle=0,width=12.5cm}
\end{center}
\caption{One gluon  exchange in the l-th power of the Wilson line. Adding up all the terms inside the square bracket gives eq. (\ref{loopvertex}) in the main text.}
\la{fig:wilsononeloopaverage}
\end{figure}

In summary:  the insertion amplitudes are
free of the non-linearity. They amount to a linear renormalization
of the C matrix, as in the second equation of (\ref{loopvertex}).
This means that they drop out when determining the value of the minimum of the effective action~\footnote{The value of the minimum is computed as an integral over C} . 
  
But they are important for the gauge independence of the effective action, and hence for the equations of motion that determine the profile, i.e. the location of the minimum.   
 
 \section{Effective potential to two-loop order and  Casimir scaling\label{sec:effaction}}

Casimir scaling has been proven at the two loop level in
a brute force way ~\cite{giovannaaltes02}~\cite{armoni}. Although not of direct relevance to our results we think it may be useful to remedy the situation.

To understand the Casimir scaling of the potential in a transparent way at the two loop level, we introduce 
 the double line representation by 't Hooft, valid   for finite N, and with a simplification for the class of models at hand. This is important because
 we will thus see that Casimir scaling is not only a consequence of 
 counting of degrees of freedom, but maybe a genuine property of  the hard modes in an interacting plasma. From the method below it will be clear that for three loops Casimir scaling necessitates constraints on planar graphs. Non-planar three-loop graphs are Casimir scaling.

  The double line representation is based on the fact that the adjoint representation can be generated  by the action on an $N\times N$ Hermitean traceless matric Q of the fundamental representation
$\Omega$ on the right and from its contragredient partner $\Omega^{-1}$ on the left:
\be
Ad\Omega(Q)=\Omega Q\Omega^{-1}.
\ee
 
Remember from previous sections %(\ref{sec:resultseffaction}) 
 the Cartan basis in group space. This basis 
was instrumental in diagonalizing the fluctuation determinant in the background of the Wilson loop. We recall its general form~\footnote{Some of the material below was developed in conversations with Rob Pisarski}.
It is a set of  $N\times N$ matrices $\l^{ij}$, where the labels i and j run both
independently from 1 to N. They are given by their matrix elements:
\be
\l^{ij}_{kl}={1\over{\sqrt 2}}\delta^{i}_k\delta^{j}_{l}.
 \la{cartanbasis}
 \ee
 
 They form an orthogonal set with norm $1/2$:
 \be
Tr\l^{ij}\l^{kl}={1\over 2}\delta^{il}\delta^{jk}.
 \la{norm}
 \ee
 
These elements form a basis for the U(N) algebra. This is obvious from the fact that i and j run independently through 1 to N. 
The N diagonal $\l^{ii}$  are an alternative to the traditional
diagonal basis:
\ba
\l^d&=&{1\over{\sqrt{2d(d+1)}}}\mbox{diag}(1,1,........,1,1-d,0,...,0), d=1,...,N-1\\  
\l^0&=&{1\over{\sqrt{2N}}}{\bf 1}.
\ea
To avoid the 
U(1) component, one has to change the diagonal members of
the Cartan basis by:
\be
\hat\l{ii}=\l^{ii}-{1\over {\sqrt{2}N}}{\bf 1} 
 \ee
 \noindent so that they become traceless.
 
 In terms of matrix elements:
 \be
 \l^{ij}_{kl}={1\over{\sqrt 2}}(\delta^{i}_{k}\delta^{j}_{l}-{1\over N}\delta^{ij}\delta_{kl}).
 \ee
 
 Now the fluctuation fields Q carry the indices $ij$ when we decompose the NxN matrix $Q=Q^{ij}\l^{ij}$.
And therefore the propagator $\<Q^{ij}Q^{nm}\> $ carries four indices
and  becomes:
\be
\<Q^{ij}Q^{nm}\> =(\delta^{im}\delta^{jn} -  {1\over N} \delta^{ij}\delta^{nm})/(p^{ij})^2.
\la{doubleprop}
 \ee
 We did not put in the Lorentz indices for notational convenience.

 In fig. (\ref{fig:doubleline}) we put the Kronecker delta's in evidence by the lines connecting the indices.
 \begin{figure}[htb]
 \begin{center}
 \psfig{file=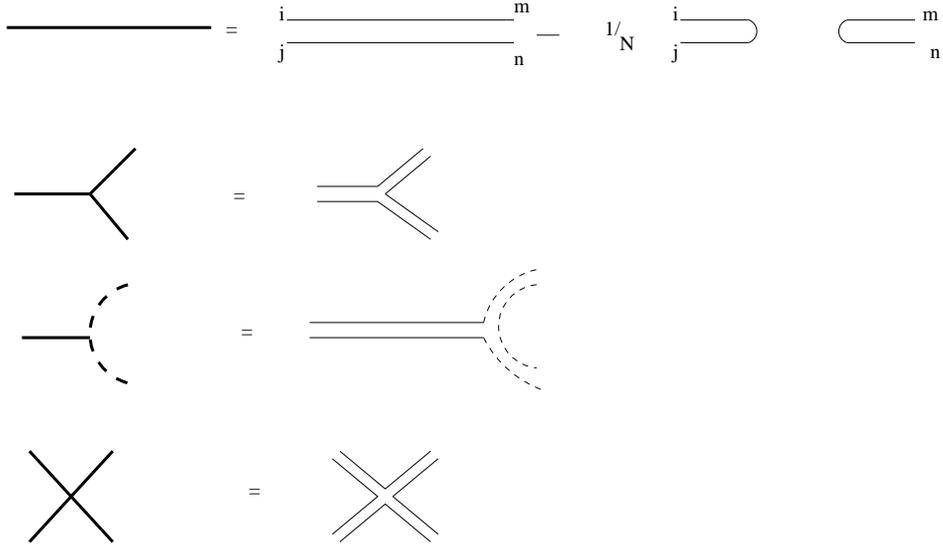,angle=0,width=12.5cm}
 \end{center}
%\centerline{\epsfig{file=doubleline.eps,width=10cm}}
\caption{Double line notation for propagators and vertices of bosons, based on (\ref{doubleprop}) and (\ref{structureconstdelta}) in the text. For fermions  and boson-fermion vertices analogous rules are valid.}
\la{fig:doubleline}
\end{figure}

The vertices in the Cartan basis are also products of Kronecker delta's. The reason is that the structure constants are formed by
commutators, and products read:
\be
\l^{ij}\l^{lm}={1\over{\sqrt{2}}}\delta^{jl}\l^{im}.
\la{products}
\ee

Together with the orthogonality relation (\ref{norm}) this gives the result:
\be
2\sqrt{2}f^{ij,lm,nr}=\delta^{ri}\delta^{mn}\delta^{jl} -\delta^{jn}\delta^{rl}\delta^{mi}.
\la{structureconstdelta}
\ee

In the figure we put only one of the two terms. The relative minus sign is important when comparing planar to non-planar diagrams.

The second term in the propagator is the "U(1) gluon", proportional in this notation to the unit matrix $\bf 1$. This gluon does not couple into any
of the vertices, since the structure constants are coming from commutators.  For the propagator of the gluino the same is valid.

Therefore in our problem we can just as well drop the 
second term in the propagator. This gives rise to a big simplification.

From the double line representation for the propagators and vertices
it follows immediately that the colour shift in the Matsubara frequencies is conserved. Take for example the three gluon vertex, and  draw a circle with midpoint in the vertex. Then the shifts in the three outgoing legs add up to:
\be
C^{ij}+C^{jl}+C^{li}=0.
\ee

Consider any one loop diagram for the free energy. We suppose it
is evaluated in  the background $C=2\pi Tq Y_k$ and we recall from section (\ref{sec:setup}), eq.(\ref{defyk}),  the form of $Y_k$:
\be
Y_k = {1 \over N} ~{\rm diag}(\underbrace{k,k,\dots,k}_{N-k~{\rm times}},
       \underbrace{k-N,k-N,\dots,k-N}_{k~{\rm times}})\nonumber. 
\la{yk}
\ee
\noindent The k are integers. The rows and columns of $Y_k$
are indexed  by i,j,...If i,j,.. are giving one of the first N-k elements
such and index is said to fall in the "k-sector" [k]. If the index gives 
one of the last elements it is said to fall into the sector [N-k].

Clearly, if the indices (ij) of a given double line are in the same sector 
  the corresponding shift in $p_0^{ij}=p_0+q((Y_k)_{ii}-(Y_k)_{jj})$
 is zero. If they are in different sectors the shift is $\pm q$. So for a one loop diagram there are 2k(N-k) possibilities of having q flowing
 in one or the opposite direction. This because there are k ways
 we can chose hte index i from the [k] sector and independently N-k ways to chose j from the [N-k] sector. The weight of any of those configurations is the same, $\sim \hat B_4(\pm q)$, and so we get 
that  sum is proportional to k(N-k) which is called Casimir scaling.

In the rest of this section we will show that any two loop diagram
will display Casimir scaling.  

For this we recall that the well-known convenience of the double line notation comes from the case where  q=0 (no background). The 
diagrams form  index cycles, and every cycle is counting for a factor
N.  This leads then to the topological classification of double line diagrams into planar, and non-planar diagrams. An example is shown in fig. (\ref{fig:2qloop}). Clearly the right most diagram has two index cycles less  than the planar ones on the left.

However if q is non-zero the counting becomes a little more complicated, and the rest of this section is handling this problem.

 \begin{figure}[htb]
 \begin{center}
 \psfig{file=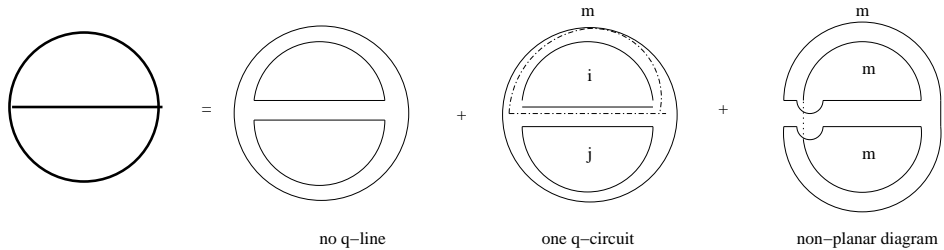,angle=0,width=12.5cm}
 \end{center}
%\centerline{\epsfig{file=2qloop.eps,width=10cm}}
\caption{A two loop diagram in double line notation.}
\la{fig:2qloop}
\end{figure}

Let us turn these simple observations into a proof that {\it any} two- loop diagram with non zero q carries a factor k(N-k).

Look at fig. (\ref{fig:2qloop}).   Any line (propagator) that carries a Matsubara frequency $(n_0+q)$ is called a q-line. So the diagram falls into two classes: one where there is 
no q-line, or one where there is at least one  q-line.
The non-planar 2 loop diagrams have only one index loop (they are $O(1/N^2)$ smaller than the planar loops with three index cycles). So they
have no q-lines.

The planar 2 loop diagrams can have q-lines as shown in fig. (\ref{fig:2qloop}).

Remember in a vertex there is conservation of colour and hence of q.
So in the two three-vertices there is an ingoing q-line and an outgoing q-line. Hence a q-line follows a closed loop, in analogy with the chemical potential for fermion lines.%~\cite{yorkkorthals}. 
An example of such a closed loop is given in fig. (\ref{fig:2qloop}). It is indicated by the broken 
line in the second diagram on the right.

We are now in a position to tell how many of those q-line diagrams
there are.  To have a q-line the two indices constituting the line must
be in different sectors of the $Y_k$ matrix. In the diagrams in fig. (\ref{fig:2qloop}) we must have m and i in different sectors. And j must be in the same sector as
m, and in a different sector from i. 

So we have the following situation in table (\ref{table2loopq}):

\begin{table}[htdp]
\caption{Counting of the multiplicity of the two-loop planar diagram. [k] is the $Y_k$ sector with N-k entries k, and [N-k] is the sector with k entries N-k. i, j and m are the indices of the index cycles in fig.(\ref{fig:2qloop})}
\begin{center}
\begin{tabular}{|c|c|}
\hline
$[k]$ & $[N-k]$\\
\hline
\hline
$mj $ & $i$\\
$i$ & $mj$\\
\hline
$j$ & $mi$\\
$mi$ & $j$\\
\hline
$m$ & $ij$\\
$ij$ &$m$ \\
\hline
$ijm$ &${}$ \\
${}$&$ijm$\\
\hline
\end{tabular}
\end{center}
\label{table2loopq}
\end{table}%

The first two lines in the table correspond to the planar diagram with the q-circuit as shown in  figure (\ref{fig:2qloop}). The next 4 lines to the two planar diagrams with the q-circuit drawn differently.
The last two lines correspond to no q-lines.

The first line has $(N-k)^2$ possibilities for the indices m and j, and 
k possibilities for the index i, so $k(N-k)^2$ in total. The second line
has $k^2(N-k)$ possibilities and describes the diagram with the same q-cycle as in the first line. So the total is:
\be
k(N-k)^2+k^2(N-k)=k(N-k)N.
\ee
The factor N is absorbed by the coupling $g^2$, and so Casimir scaling is verified! 

For q=0 we should find the familiar factor $N(N^2-1)$ for the free energy.  The terms just discussed contribute $3k(N-k)N$.The last two lines in the table contribute $k^3+(N-k)^3$. The contribution of the non-planar graph comes  with a relative minus sign~\footnote{Vertices in non-planar graphs get their contributions from terms
 in the commutators which have opposite sign} and equals
$-(k+(N-k))$, because there is only one q circuit. The sum is indeed  the familiar free energy factor.

From this  analysis one can draw an immediate conclusion for the 3 loop potential: the non-planar contributions with  only two (or less) q cycles will show Casimir scaling.  But the planar diagrams 
necessitate relations between weights
.

 % \begin{figure}
%\begin{center}
%\psfig{file=wilsononeloopaveragesusy.eps,angle=0,width=12.5cm}
%\end{center}
%\caption{One gluon  exchange in the l-th power of the Wilson line. Adding up all the terms inside the square bracket gives eq. (\ref{wilsonloopav}) in the main text. }
%\la{fig:wilsononeloopaverage}
%\end{figure}
% 
 
 \section{Minimizing the effective action\la{sec:minimization}}

From the explicit expressions for kinetic and potential terms in the 
effective potential developed in the previous sections we now 
find the mimimum of the effective potential by varying the profile $q$:
\be
U=\int dx (K+V).
\ee

The kinetic term is proportional to $q'^2$, $K=\widehat K q'^2$. Then the minimum of U is found by quadrature:
\be
U_{min}=\int dx\bigg(\sqrt{\widehat K} q'-\sqrt{V}\bigg)^2+2\int_0^1 dq\sqrt{\hat K V}.
\la{minu}
\ee

The first term is non-negative definite and produces the equations of motion for the profile $q$ by setting it zero. The second term is the value of the minimum. 

There are two poles  in the kinetic term. One is at q=0 and is canceling with the corresponding zero of $\sqrt{V}\sim q(1+\mbox{regular in q})$. But the pole at $q=1/2$ in $\psi(1/2-q)$ causes a logarithmic divergence. This pole is due to a single Matsubara frequency in the sum for the gluinos. In fact the denominator
$|n+{1\over 2}+q(x)|$ in eq. (\ref{kinetic}) becomes bosonic at the center of the wall,
where $q(0)=1/2$.  So the gluino becomes at the very center of the wall a boson.

And its Matsubara frequency  $n=-1$ causes the divergence. We will just by fiat subtract it  so replace $\psi(1/2-q)$ by
\be
\psi(1/2-q)-{1\over {1/2-q}}.
\la{polesubtract}
\ee 

The divergence  has an interesting physical explanation, and we come back to 
it in section \ref{sec:discussion}. %Our ad hoc subtraction of it will be justified.

 To wet the appetite of the reader: the remedy is {\it not} simply to reinstall the thermal
mass of the gluino in the subtracted pole in eq. (\ref{polesubtract}). It would  help to cure the divergence because
\be
\int^{1/2} dq{1\over{1/2-q+m_\l(q=1/2)}}\sim\log(m(q=1/2)).
\la{cure}
\ee
 However, the thermal gluino mass turns out to be zero for $q=1/2$!

To get the tension  we plug  in the results for K and V from eq. (\ref{renkinterm}) and (\ref{effpot}), and using (\ref{polesubtract}) one finds:
\ba
{\r_k(T)\over{k(N-k)T^2}}&=&{2\sqrt{2}\pi^2\over 15}{(9-2\sqrt{3})\over{\tilde g}}\D^{1\over 2}\bigg[1-{\tilde g^2\over{(4\pi)^2}}\bigg\{\{(-2.92683 ...)\times b_b(\d)\nonumber\\
&+&(3.27471...)\times b_f(\d) +(1.96351..)\times {b_0(\d)\over 2}\}+ ({\d-2\over 2})\times 5\bigg\}\bigg]
\la{rhoresult}
\ea
For gluodynamics we have:
\ba
\r_k&=&k(N-k){4\pi^2T^2\over{3\sqrt{3\tilde g^2}}}\times\nonumber\\
&&\bigg[1
-{\tilde g^2\over{(4\pi)^2}}\bigg\{\{(-2.98505...+1.96351..)\times (-{11\over 3})-{1\over 3} -2\}+{1\over 2}\times 5\bigg\}\bigg]
\la{tensiongluo}
\ea

 The terms within small curly brackets stem from the kinetic tem folded with  the square root of the lowest order potential.
 
Numerical values for $\psi(1/2)=-1.96351..$~\cite{rg} and 
for the folding of  the psi-functions with the square root of the 
lowest order potential are used. 

\section{Discussion\la{sec:discussion}}

First we justify dropping the gluino mode that gets bosonic in the center of the wall.  

To this end we recall the third guise for $\r_k$ as the energy
of  a one dimensional soliton as explained at the end of  section \ref{sec:setup}. The soliton is  stretched along the y-direction, at x=0. It separates two regions in the x-y plane
with the Polyakov loop having the value 1, respectively $\exp(ik{2\pi\over N})$.

The reason we can drop the mode that diverges at x=0  is that it 
has zero energy in the field of the soliton. Remember that the  time is the z-direction, so zero-energy implies for the gluino spinor:
\be
\G_z\pa_z\l=0.
\ee

To find the zero-energy solution, write the Dirac equation for the gluino, with the momentum in the $\t$ direction $\pi T-2\pi Tq(x)$:. this is the mode causing the divergence in $\psi(1/2-q)$ for $q=1/2$ in the effective kinetic term, eq. (\ref{renkinterm}).

We have only  x  dependence left for the zero-energy mode:
\be
\bigg(\G_x\pa_x+\G_0(\pi T-2\pi Tq(x))\bigg)\l(x)=0
\la{diraczeromode}
\ee

So the solution is~\cite{jackiwrebbi1976}, u being a constant spinor:
\be
\l_0(x)=\exp\bigg(\G_0\G_x\int_0^xdx'(\pi T-2\pi Tq(x'))\bigg)u.\la{diracsolution}
\ee
To have a localized  state $\l_0$ we need $\G_0\G_xu=u$. 

We have found that the diverging mode is a zero mode. It binds to the soliton and forms a state degenerate with it. Hence it is legal to drop it.

There are several interesting aspects of this state that will be discussed in a sequel paper.

 \begin{figure}[htb]
 \begin{center}
 \psfig{file=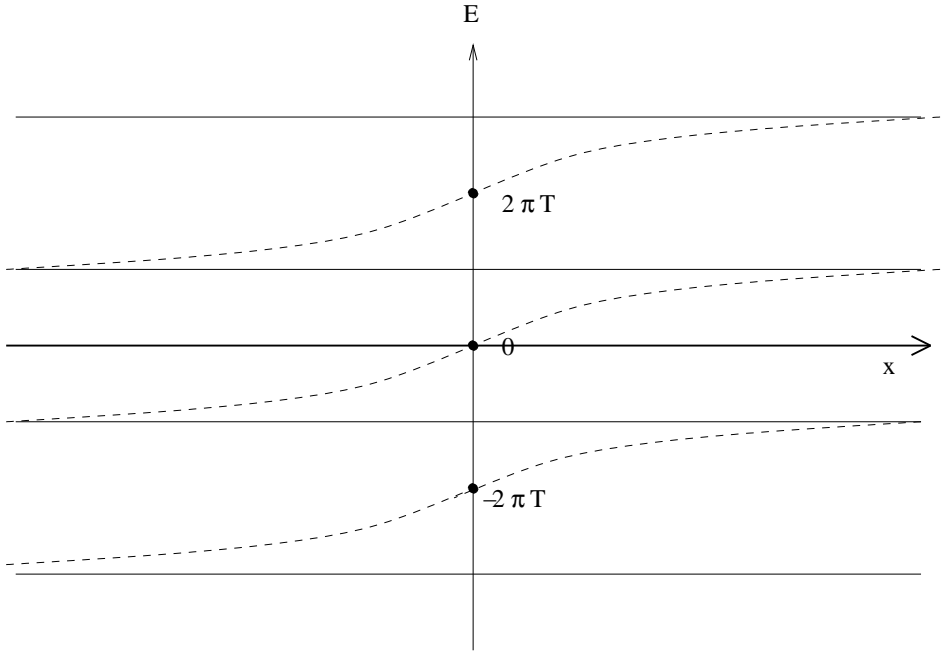,angle=0,width=12.5cm}
 \end{center}
%\centerline{\epsfig{file=2qloop.eps,width=10cm}}
\caption{Behaviour of the half-integer gluino energies due the presence of a magnetic flux. Vertical axis is energy in units of $2\pi T$, horizontal axis is the x-axis, orthogonal to the magnetic flux in the y-direction at x=0 (see fig. (\ref{fig:guises}~.b). Gluinos are anti-periodic in the short spatial direction $\t$, and have zero momentum in x and y direction, as in eq.(\ref{diraczeromode}). The continuity of the derivatives at the jump at x=0 is manifest. }
\la{fig:levelcrossing}
\end{figure}

 We mention as a curiosity the contribution from the potential to the tension: for gluodynamics it is from 
eq. (\ref{tensiongluo}) $5/2$, and for ${\cal N}=1, 2, 4$ it is given by ${\d-2\over 2}5$. 
In the case of gluodynamics this is already true on the level
of the effective potential, eq. (\ref{gluodynpot}), for any value of q as in eq. (\ref{gluodynpot}). In the case of SUSY it is only true for the minimum $\r_k$ , i.e. only after integration over q.

The kinetic part is subject to ambiguity due to the choice of the running coupling. With our choice
the  $\OO(g^2)$ correction is negative and qualitatively like in gluodynamics.

\subsection{The profile to lowest order and the fluctuation operator}

Up till now we did not look at the other piece of information the 
effective potential holds in store: the profile corresponding to the tension. 
 
 The profile follows from the first term in eq.  (\ref{minu}), substituting (\ref{effpot}), (\ref{difference}) and 
 (\ref{bernoulli2}) and (\ref{bernoulli3}).
 The equation of motion becomes to lowest order in terms of q and $\xi=m_Dx$, $m_D$ the Debye mass from eq. (\ref{screeningmassgauge}):
 \be
 q'=q \sqrt{1-{4q\over 3}}
 \la{profileeqmotion}
 \ee
 
 Note that the profile does not depend on k, the strength of the loop, and that this equation is valid for $0<q<1/2$. Continuity of the kinetic term and symmetry arguments  extend it to the upper half of the interval. Clearly
 the profile is for small q exponentially decreasing as $\exp(-m_D|x|)$. 
 
 The solution for the profile is:
 \be
 q_m={3\over 4}(1-\tanh^2(-3m_Dx/8 +x_0))
 \la{profile}
 \ee
\noindent with $x_0$ such that $q(0)=1/2$.
 
 The $\OO(g^2)$ correction to the profile is straightforward as long as we stay away from its wings (small q). The reason is the pole at q=0 in the kinetic term. This pole signals the well-known presence of corrections
 to the Debye mass  which are not $\OO(g^2)$, but larger, by a logarithmic amount.  In ref.~\cite{giovannaaltes02} this problem is discussed in more detail.
 
A  last remark concerns the $\OO(g^3)$ corrections. They can be obtained from their gluodynamic analogue, as the following argument shows.

 The second derivative of the lowest order potential $V_1$. It reads:
\be
V_1''(q)=m_D^2(1-4q).
\ee
 
 This turns out to be the one loop self-mass of the time component of a stationary, zero spatial momentum  gluon, in the background q.
 %_kSuch a gluon is either colour diagonal or off-diagonal
% but with its indices in the same sector of the hypercharge $Y$.
The gluon itself is supposed to be q-neutral, so is not screened by the background. Such gluons are the cause of the familiar, electric infra-red divergencies present in the plasma without background.

 The q-independent part of the self-mass gives upon resummation the Debye mass
 corrected gluon propagator, and hence the $OO(g^3)$ correction to the free energy:
 \be
\mbox{Free energy due to soft modes}\sim1/2\log \det (-\nabla^2+m_D^2)\sim m_D^3\sim g^3.
 \ee
 
 Taking into account the q dependence of the  selfmass one computes the same correction but now to the tension  $\r_k$ through the determinant. Since we now compute soft, $\OO(gT)$,
 fluctuations, which are of the same order as the length scale in the  profile  the use of gradient expansions is not allowed in this determinant. So one has to evaluate the eigenvalue spectrum 
 of the  Schroedinger equation:
 \be
 \mbox{tension due to soft modes}\sim{1\over 2}\log
\det \bigg(-\nabla^2+ m_D^2(1-3/\cosh^2(-3m_Dx/8+x_0))\bigg).
 \la{schroe}
 \ee
 However this calculation has already been done in gluodynamics\cite{giovannaaltes02}, where the fluctuation operator $V''_1(q_m)$ is identical, up to some trivial rescalings.

%\pagebreak

\section{Conclusions \la{sec:conclu}}

In this paper the corrections to the tension of the 't Hooft loop for hot SUSY theories have been computed to two loop order. The cubic order  is shown to be a fairly trivial rescaling of the same 
problem in gluodynamics. The calculation can be done analytically because the one-dimensional Schroedinger problem, eq. (\ref{schroe}),  is separable.

  What can the understanding of the spatial 't Hooft loop bring us in terms of a better understanding of the hot phase? After all, the most we learn from it seems to be that it scales according to the
  Casimir scaling law, $k(N-k)$ and that this scaling law is characteristic for gluons, the screened quasi-particles,  being in the adjoint representation. This comes hardly as a surprise.%! It seems we are shooting with a gun at a mouse.
 
 However, in ${\cal N}=4$  theory the self-duality of the theory at
 zero temperature may help to shed light on the behaviour of the 
 spatial Wilson loop. The latter measures the mean flux of magnetic quasi-particles . It is well-known from lattice calculations~\cite{teper} that also the Wilson loops show Casimir scaling. It is tempting to associate to this scaling magnetic quasi-particles in the adjoint representation in some magnetic group~\cite{giovannaaltes01}\cite{korthalsmeyer}\cite{korthalsschools} and ${\cal N}=4$ theory seems a setting where one may have a more quantitative grip on these ideas. 
% To take an amusing example, the spatial Wilson loop in the  defining representation  has in strong coupling ($\tilde g>>1$, but $g<<1$) and large N
% the value~\cite{divecchia}:
% \be
% \s={\pi\over 2}\sqrt{\tilde g^2}T^2. 
% \ee
% Changing $g^2/4\pi$ into $4\pi/g^2$ and $\vec B $ into $-\vec E$
% we get the value of the 't Hooft loop in the defining representation:
% \be
% \r_1=2\pi^2 N/\sqrt{\tilde g^2}.
% \ee
% Comparing to the value for $\r_1$   in ${\cal N}=4$, eq. (\ref{rhoresult}) with $\d=10$, there is a factor 1.04386, 4\% off.  This may well be fortuitous, because both results are obtained in weak gauge coupling, which is excluded by the dual couplings.

\section{Acknowledgements}
The author thanks Kareljan Schoutens for kindly extending hospitality at  the Institute for Theoretical Physics of the University of Amsterdam, and for an inspiring atmosphere,  Adi Armoni, Prem Kumar and Jefferson  Ridgway  for bringing the interest of computing higher order corrections to his attention and for instructive e-mail exchange. He is indebted to  Rob Pisarski 
for his input concerning the double line representation for any finite number of colours. Laurent Lellouch and Vincent Bayle graciously provided 
the author with computational means. Discussions with Misha Stephanov and Dam Son about the leading term for the tension in 2002 at KITP,  Santa Barbara, were quite useful. Incisive remarks by Kostas Bachas and his hospitality at the ENS, Paris, helped understanding the   gluino zero-mode problem.

\section*{Appendix A: Determinant in mixed x and p representation}
In this appendix we elaborate on how to compute the 
determinant of the d'Alembertian with a background B,
where the variation with x does not commute with the corresponding derivative. We suppose the colour degrees of freedom have been diagonalized in the Cartan basis $Q^{ij}$, so that the background $B$ below stands for $B^{ij}=B_{ii}-B_{jj}$.

We start with eq. (\ref{trlogx}) in the main text and write
\be
\Tr \log(- D^2(\underline{B}))-\Tr\log(- D^2(B))
\la{trlogx1}
\ee
\noindent for the difference between the non-commuting
(underlined background B) and the commuting case ( written as B).

We write:
\be
\underline{B}=B(X).
\ee
The operator X and the momentum operator in the x-direction have the commutation relation as in eq. (\ref{pxrelations})
\be
[X,P_x]=-i
\ee 

\noindent and they act in a Hilberspace spanned by the normalized kets $\vert x\rangle$, with $X\vert x\rangle=x\vert x\rangle$.
As in the main text we introduce the eigentime t and write 
the difference of d'Alembertians as:
\be
-\Tr\int{dt\over t}\bigg [\exp(-t(-D(\underline{B}^2))-\exp(-t(-D(B)^2)\bigg].
\label{eigentime}
\ee

The trace becomes in this language ($l_0=2\pi T n$, n integer)
\ba
&&\Tr ~\exp\bigg(-t ((l_0+B)^2+P_{tr}^2+P_x^2)\bigg) \nonumber\\
&=&T\sum_{n}\int {dl_{tr}^{d-2}\over{(2\pi)^{d-2}}}\mu^{4-d}\int dx\langle x\vert\exp\bigg(-t ((l_0+B(X))^2+l_{tr}^2+P_x^2)\bigg) \vert x\rangle.
\label{traceexplicitlogx}
\ea

The background field $B(X)=B(x+(X-x))$ is expanded 
around $B(x)$, the field that commutes with $P_x$:
\be
B(X)=B(x)+B'(x)(X-x)+{1\over 2}B''(x)(X-x)^2+...
\ee
We know already- from the absence of a classical potential term- that every derivative $B'(x)={dB(x)\over{dx}}$ counts as
a factor $g$, and since we are interested in order $g^2$ we 
neglect the dots.

The matrix element in (\ref{traceexplicitlogx}) becomes on substitution of the expansion for $B(X)$ ( we use the notation $2A=((l_0+B(x)))^2)'$ and $\Omega=A'$:
\ba
&~&\langle x\vert\exp(-t (l_0+B(X))^2+l_{tr}^2+P_x^2)) \vert x\rangle\nonumber\\
&=& \langle x\vert\exp(-t (l_0+B(x))^2 +\Omega(X-x-{A\over{\Omega^2}})^2-{A^2\over{\Omega^2}}+l_{tr}^2+P_x^2)) \vert x\rangle\nonumber\\
&=&\langle {A\over{\Omega^2}}\vert\exp(-t (l_0+B(x))^2+\Omega(X
)^2-{A^2\over{\Omega^2}}+l_{tr}^2+P_x^2)) \vert  {A\over{\Omega^2}}\rangle. 
\la{matrixelement}
\ea
The dependence on x in the last expression has been translated away. It stays only implicitely in $B(x)$. 

In what follows we assume it makes sense to expand the 
integral  (\ref{traceexplicitlogx}) in terms of A  and $\Omega$. 

We note that in case $\Omega=0$ the matrix element becomes proportional to:
 \be
 \langle x\vert\exp(-tP_x^2)\vert x\rangle={1\over{(4\pi t)^{1\over 2}}}.
 \ee
 This is precisely the result one would get from the $l_x$ integration, pretending the background is constant.
 
In our case however we have $\Omega={\cal O}(g^2)$ and borrow a formula from \cite{itzyk} to take its presence into account:
\be
\langle{A\over{\Omega^2}} \vert|\exp(-t(\Omega X^2+ P_x^2))\vert {A\over{\Omega^2}}\rangle
 ={a^{1\over 2}\over{(4\pi t)^{1\over 2}}}\exp\bigg(-{a\over{2t}}{A^2\over{\Omega^4}}(-1+\cosh\omega)\bigg)
\la{oscillatorr}
\ee
\noindent with
\ba
a&=&{\omega\over{\sinh\omega}}\\
\omega&=&2\Omega t.
\label{oscillator}
\ea

In the second term expand a and $\cosh\omega$ for small $\omega$
and  substitute the result in (\ref{matrixelement}).
Terms with $\Omega$ can  be  transformed	
\be
\int dx \exp(-t(l_0+B(x))^2)\Omega=\int \exp(-t(l_0+B(x))^2)2A^2t
\ee
through partial integration in x and the definition of A and $
\Omega$ above eq. (\ref{matrixelement}). The boundary terms at $x=\pm\infty$ cancel with each other for $B(\pm \infty)=0, ~\mbox{or}~2\pi T$ after summing over all n in $l_0=2\pi nT$. The summation over Matsubara modes introduces the periodicity in B mod $2\pi T$, hence the cancellation.

The result of this expansion becomes for the matrix element in (\ref{oscillator})  simply:
\be
 \langle{A\over{\Omega^2}} \vert\exp(-t(\Omega X^2+ P_x^2))\vert {A\over{\Omega^2}}\rangle={1\over{(4\pi t)^{1\over 2}}}
(1-{1\over 3}A^2t^3+..).
\la{finalexp}
\ee

Looking back at eq. (\ref{traceexplicitlogx}) we see that the integrations over $l_{tr}$ produce a power of eigentime $T^{{2-d\over 2}}$. Together with the powers of t coming from the expansion in the oscillator frequency $\Omega$ above , i.e. $t^{3\over 2}$,  one arrives at a power $t^{5-d\over 2}$.

This power ensures that the term with the factor $A^2=(l_0+B(x))^2B'^2$ becomes,
after substitution into (\ref{eigentime}), through partial integration in the eigentime t:
\be
A^2={(5-d)\over {2t}}B'^2,
\la{5-d}
\ee  
\noindent hence our final result for (\ref{trlogx1}) becomes, using (\ref{eigentime}), (\ref{finalexp})  and (\ref{5-d}) :
\ba
&&\Tr \log(- D^2(\underline{B}))-\Tr\log(- D^2(B))\nonumber\\
&=&{1\over 3}\int dx ~B'^2{1\over{(4\pi)^2}}\bigg({T\over{\mu}}\bigg)^{-2\epsilon}{5-d\over 2}{\Gamma({5-d\over 2})\over{\pi^{{5-d\over 2}}}}\sum_n{1\over{|n+q|^{5-d}}}.
\la{resultapp}
\ea

We wrote $q=B(x)/2\pi T$, and $\epsilon=2-d/2$, and to complete the result we take from \cite{rg} for $d\rightarrow 4$:
\be
\sum_n{1\over{|n+q|^{5-d}}}={1\over{\epsilon}}-\psi(q)-\psi(1-q).
\ee
The $\psi$ function is the logarithmic derivative of the Euler $\Gamma$ function. It has a simple pole at $q=0$.

In the main text  ( eq. (\ref{trlogx}) and (\ref{trlogxf}) ) expressions of the form:
\be
\Tr\log(-D^2(\underline B)\pm B')-\Tr\log(-D(B))
\ee
\noindent appear.

Expanding the  B' gives us for half  the sum of the terms with
plus and minus sign:
\be
\int dx{ B'^2\over 2} \Tr{-1\over{(-D(B))^2}}+\Tr\log(-D(\underline B)^2)-\Tr\log(-D(B)^2
\la{sumpm}
\ee

By the same technique the first term leads to
eq. (\ref{resultapp}), with the factor ${1\over 3}{5-d\over 2}$ replaced 
by $-{1\over 2}$.

The  difference between boson respectively  fermion trace is the periodicity respectively the anti-periodicity. Hence it comes into
play in the argument of the $\psi$ functions. For bosons
the argument is $q$, for fermions $q+{1\over 2}$.

The former produce a pole at $q=0$, the latter  a pole at $q=1/2$, as discussed in the main text. Both these poles signal the 
presence of infra-red effects, but are remedied in different ways.

\section*{Appendix B: Bernoulli polynomials}

They are related to the polynomials defined in (\ref{bernoullihat}):
\ba
\hat B_{d=4}(x)&=&{2\over 3}\pi^2T^4 B_4(x)\\
\hat B_3(x)&=&{2\over 3}\pi T^3 B_3(x)\\
\hat B_2(x)&=&{1\over 2}T^2 B_2(x)\\
\hat B_1(x)&=&-{T\over{4\pi}} B_1(x)
\la{bernoulli2}
\ea
\noindent and finally ($\e(x)$ is the sign function):
\ba
B_4(x)&=&x^2(1-|x|)^2\\
B_3(x)&=&x^3-{3\over 2}\e(x)x^2+{1\over 2}x\\
B_2(x)&=&x^2-\e(x)x+{1\over 6}\\
B_1(x)&=&x-{1\over 2}\e(x)
\la{bernoulli3}
\ea
\section*{Appendix C: Spinor projectors in $\d$ dimensions}
We establish the form of projectors on Majorana-Weyl spinors in 10
dimensions, on Weyl spinors in  6 dimensions and Majorana spinors in 4 dimensions. This  is important for establishing the 
dimensional factors in the fermion amplitudes.\\
$\bullet$$\d=10$. The projector ${\bf 1}_f=\l\bar\l$ is defined by $\bar\l=\l^T{\bf C}$ and $\G_{11}\l=\l$.\\
$\bullet$ $\d=6$.The projector is defined by $\G_7\l=\l$\\
$\bullet$ $\d=4$. The projector is defined by $\bar\l=\l^T{\bf C}$\\
In all these cases one has $\Tr \G_{0x}{\bf 1}_f=\d-2$.

\section*{Appendix D: Individual diagrams contributing to the potential}

We compute the potential term $V=V_1+\tilde g^2V_2$ in the effective action:
\be
U=\int dx\bigg[K(q) q'^2+V(q)\bigg].
\ee
We refer to fig. (\ref{fig:susyonetwolooppot}) and compute in Feynman gauge. All two-loop results below are given in terms og the 't Hooft coupling $\tilde g^2=g^2N$. The hatted Bernoulli
polynomials are given in Appendix B.

 We get  for the bosonic graphs (a):
 \ba 
a_1&=& {\d\over 2}2k(N-k)\hat B_4(q)\nonumber\\
a_2&=&-{3\over 4}(\d-1)k(N-k)\bigg(\hat B_2(q)^2+2\hat B_2(q)\hat B_2(0)\bigg) \nonumber\\
a_3&=&{1\over 4}\d(\d-1)k(N-k)\bigg(\hat B_2(q)^2+2\hat B_2(q)\hat B_2(0)\bigg) .
\la{bosonic}
\ea
For the ghost graphs (c):
\ba
c_1&=&-2k(N-k)\hat B_4(q)\nonumber\\
c_2&=&{1\over 4}k(N-k)\bigg(\hat B_2(q)^2+2\hat B_2(q)\hat B_2(0)\bigg)  .
\la{ghost}
\ea

For the  one loop fermion and mixed boson-fermion graphs ($q_f=q+1/2$):
\ba
b_1&=&-{\d-2\over 2}2k(N-k)\hat B_4(q_f)\nonumber\\
b_2&=&\bigg({\d-2\over 2}\bigg)^2k(N-k)\bigg[\hat B_2(q_f)^2+2\hat B_2(1/2)\hat B_2(q_f)\nonumber\\
&-&2\bigg(\hat B_2(q_f)\hat B_2(q)+\hat B_2(1/2)\hat B_2(q)+\hat B_2(0)\hat B_2(q_f)\bigg)\bigg]
\la{fermionic}
\ea

Compare  to eq. (\ref{effpot}) and (\ref{difference}) in the main text.
Adding $a_1, b_1$ and $c_1$ gives $V_1$ . $V_2$ is obtained
by adding $a_2, a_3,  b_2$ and $c_2$.

\end{document}